\documentclass[
    final             % use final for the camera ready runs
    ,numberedheadings % uncomment this option for numbered sections
  ]
  {aipproc}

\layoutstyle{8x11single}

\def\fd{\hbox{$.\!\!^{\rm d}$}}

\def\degr{\hbox{$^\circ$}}
\def\sun{\hbox{$\odot$}}

\begin{document}

\title{Symbiotic Stars with Similar Line Profiles during Activity\,\thanks{Based on observations collected at the Rozhen National Astronomical Observatory, Bulgaria}}


\author{N.~A.~Tomov}{
  address={Institute of Astronomy and National Astronomical Observatory, Bulgarian Academy of Science,\\ PO Box 136, BG-4700 Smolyan, Bulgaria\\}
}

\author{M.~T.~Tomova}{
  address={Institute of Astronomy and National Astronomical Observatory, Bulgarian Academy of Science,\\ PO Box 136, BG-4700 Smolyan, Bulgaria\\}
}

\author{D.~V.~Bisikalo}{
  address={Institute of Astronomy, Russian Academy of Science, 48 Pyatnitskaya Street, RU-119017 Moscow, Russia}
}

\begin{abstract}
Line profiles containing indication of bipolar collimated outflow along with P~Cyg absorption during phase of activity of several symbiotic systems are considered. The H$\gamma$ profile of Z~And during its 2006 outburst consisted of four groups of components.
The profile of the \mbox{He\,{\sc i}} $\lambda$\,5876 line of the Hen~3-1341 system during its 1998--2004 outburst had
high-velocity satellite components on one hand and a broad P~Cyg absorption on other hand. The H$\alpha$ and H$\beta$ profiles of the BF~Cyg system during its 2006--2012 outburst had satellite components, observed for the first time, along with P~Cyg absorptions. These profiles are interpreted in the light of a model related to the strong recurrent outbursts of Z~And. The AG~Dra system during its 2006 outburst did not contain indication of collimated outflow. Its profiles are interpreted in the light of the model related to the first outburst of 2000--2012 active phase of Z~And.
\end{abstract}

\maketitle

%%%%%%%%%%%%%%%%%%%%%%%%%%%%%%%%%%%%%%%%%%%%
%% MAINMATTER
%%%%%%%%%%%%%%%%%%%%%%%%%%%%%%%%%%%%%%%%%%%%

\section{Introduction}
\label{sec:intro}

Symbiotic stars are long-period interacting binaries consisting of a cool visual primary and a hot compact secondary component accreting matter from the atmosphere of its companion. Their spectral variability is determined from both the orbital motion and the outburst events of the hot component which are often accompanied by intensive loss of mass in the form of optically thick shells, stellar wind outflow and bipolar collimated jets.  The nature of the collimated jets is subject of intensive theoretical investigation and the view they represent outflow from an accreting compact object is widely accepted \citep{Zanni+07}. The collimated bipolar outflow, however, could arise due to a stellar wind, if mechanism of collimation is available in the system. Such a mechanism can be related to disc-like formation surrounding the white dwarf which provides a small opening angle of the outflowing jets.

The Z~And system is considered as a prototype of the classical symbiotic stars. It consists of a normal cool giant of spectral
type  M4.5 \citep{MS}, a hot compact object with temperature higher than 10$^5$~K \citep{FC88,Sok06} and an extended surrounding nebula partly photoionized by the hot component. Its orbital period is 758\fd8, which is based on both photometric \citep{FL94} and radial velocity \citep{MK96} data.
The last active phase of Z~And began at the end of 2000 August \citep{Sk00} and continues up to now including  seven optical eruptions. The maxima of the light during this phase were in 2000 December, 2002 November, 2004 September, 2006 July, 2008 January, 2010 January and 2012 January \citep[AAVSO data]{Sk09}. A scenario for interpretation of the line spectrum of Z~And during this active phase containing two stages of the evolution of the accreting compact object was suggested in the works of  \citet{T1+10,T+11}. It is supposed that a thin accretion disc from accretion of a stellar wind exists in the system and the observed two velocity regime of mass outflow from the compact object during the first outburst of the active phase is explained with a collision of one part of the outflowing material with the disc and decrease of its velocity. The first outburst (and every following one) gives birth to conditions a disc-like envelope surrounding the accretion disc to form in the system, which can collimate the outflowing material during the following outbursts and the  growth of the collimated outflow in parallel with the development of the wind is thus explained. During the first outburst of Z~And in 2000--2002 two velocity regime of the mass outflow was observed. During its 2006--2007 outburst development of a stellar wind together with a bipolar collimated outflow was observed on the other hand. Some other classical symbiotic stars have similar behavior during their active phases. The mean aim of this work is selected lines of some other symbiotic binaries during their active phases to be considered and to show that their profiles can be interpreted in the framework of the scenario proposed for Z~And.

\section[]{Observations and reduction}
\label{sec:obs}

The spectroscopic observations were obtained with the Photometrics CCD camera mounted on the Coud\'e spectrograph of the 2m Ritchey--Chr\'etien-Coud\'e (RCC) telescope of the Rozhen National Astronomical Observatory. The spectral resolution was
$0.2$~\AA\,px$^{-1}$ on all occasions. When more than one exposure was taken per night, the spectra were added with the aim of improving the signal-to-noise ratio. The IRAF package\footnote {The IRAF package is distributed by the National Optical Astronomy Observatories, which is operated by the Association of Universities for Research in Astronomy, Inc., under contract with the National Science Foundation.} was used for the data reduction as well as for obtaining the dispersion curve, calculating the radial velocities and equivalent widths.

The line fluxes were dereddened on using the extinction law in the paper by \citet{CCM89}.

\section[]{The spectral variability of Z~And}
\label{sec:zand}

\subsection{The model}
\label{subsec:model}

We proposed a model for interpretation of the line spectrum of Z~And during its 2000--2012 active phase \citep{T1+10,T+11}. It was concluded that as a result of accretion of the stellar wind from the giant star a thin accretion disk, located in the orbital plane, forms around the compact object in the quiescent state of the system. The estimates of the size and mass of the disc show its innermost part can be optically thick in the quiescent state. During the active phase an accretion disc with a mass of $50$--$80$ per cent of its initial mass exists, too (see \cite{T1+10} and references therein).

\begin{figure}
	\includegraphics[width=0.47\textwidth]{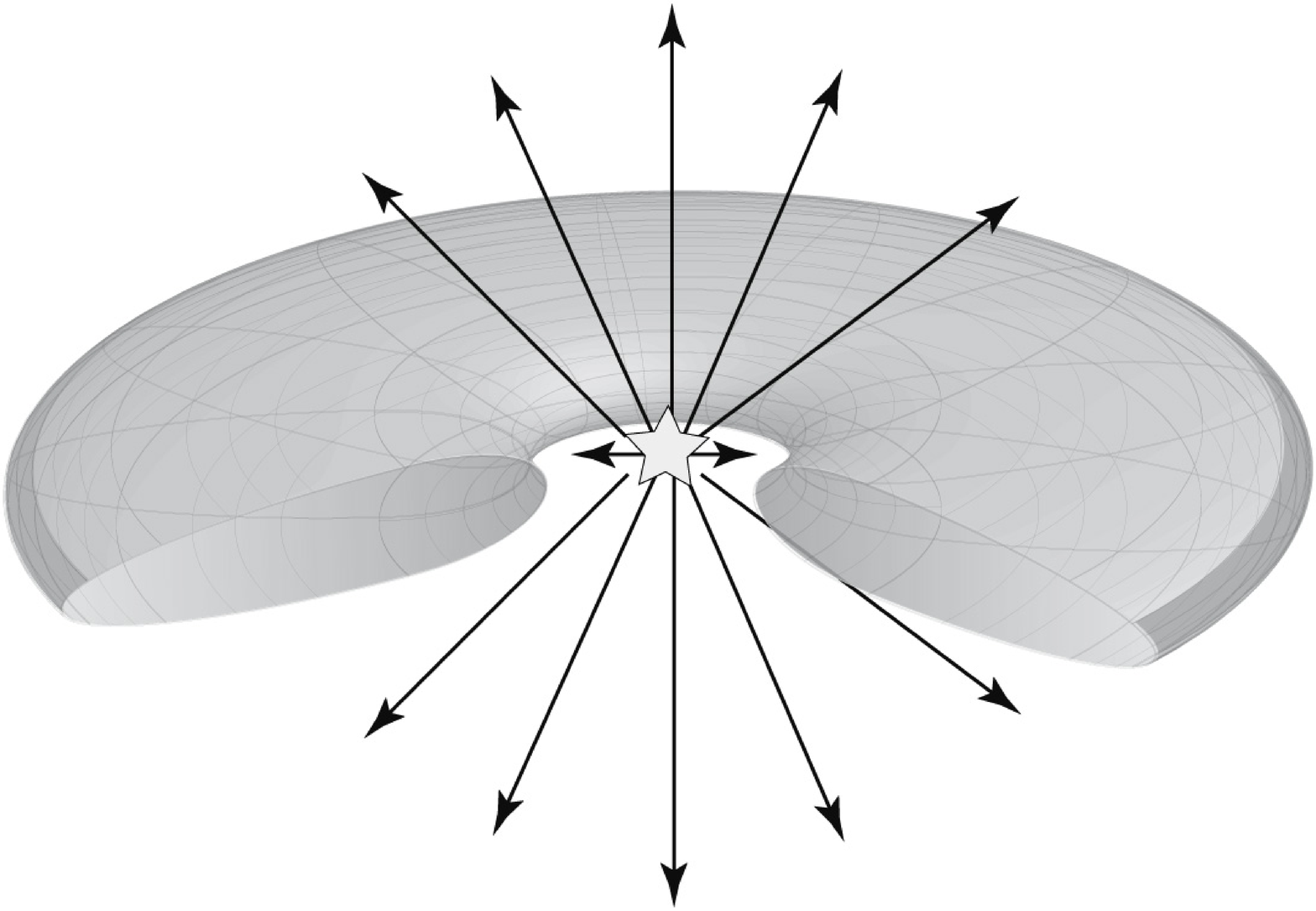}
	\includegraphics[width=0.47\textwidth]{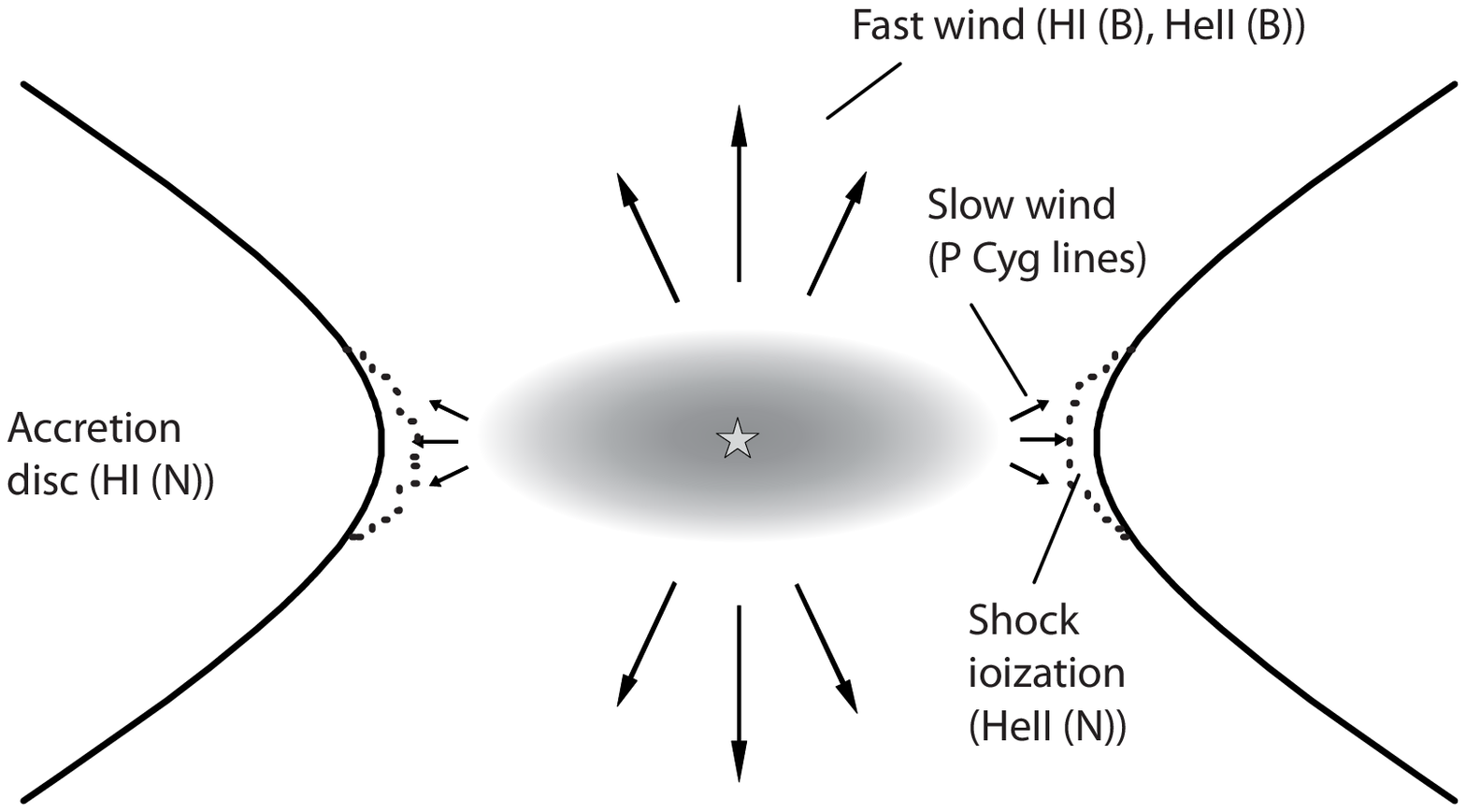}
\caption{\emph{Left panel}: Schematic model of the region around the hot component during the first outburst.
	\emph{Right panel}: The same, but in the plane perpendicular to the orbital plane where the emission regions are shown. (From spectroscopy presented in \citet{TTB08}.)}
	\label{model1}
\end{figure}

During the first outburst the outflowing material with a high velocity collides with the accretion disc. As a result, the ejecta velocity decreases in the region of the orbital plane and does not change at higher stellar latitudes (Fig.~\ref{model1}, left panel). The decrease of the velocity leads to an increase of the density and the level of the observed photosphere resides further away from the star. At higher stellar latitudes the level of the photosphere resides closer to the star. In this way an optically thick disc-like shell forms in the orbital plane, which plays the role of the observed photosphere. This shell occults the hot compact object and since the shell has a lower effective temperature \citep{Sk09,TTT03,Sk06} it is responsible for redistribution of the continuum energy and a growth of the optical flux of the star. The observed P~Cyg absorption is related to this shell (see Fig.~\ref{model1}, right panel).

%%%%%%%%%%%%%%%%%%%%%%%%%%%%%%  FIGURE 1 model1

\begin{figure}%[!htb]
	\includegraphics[width=0.47\textwidth]{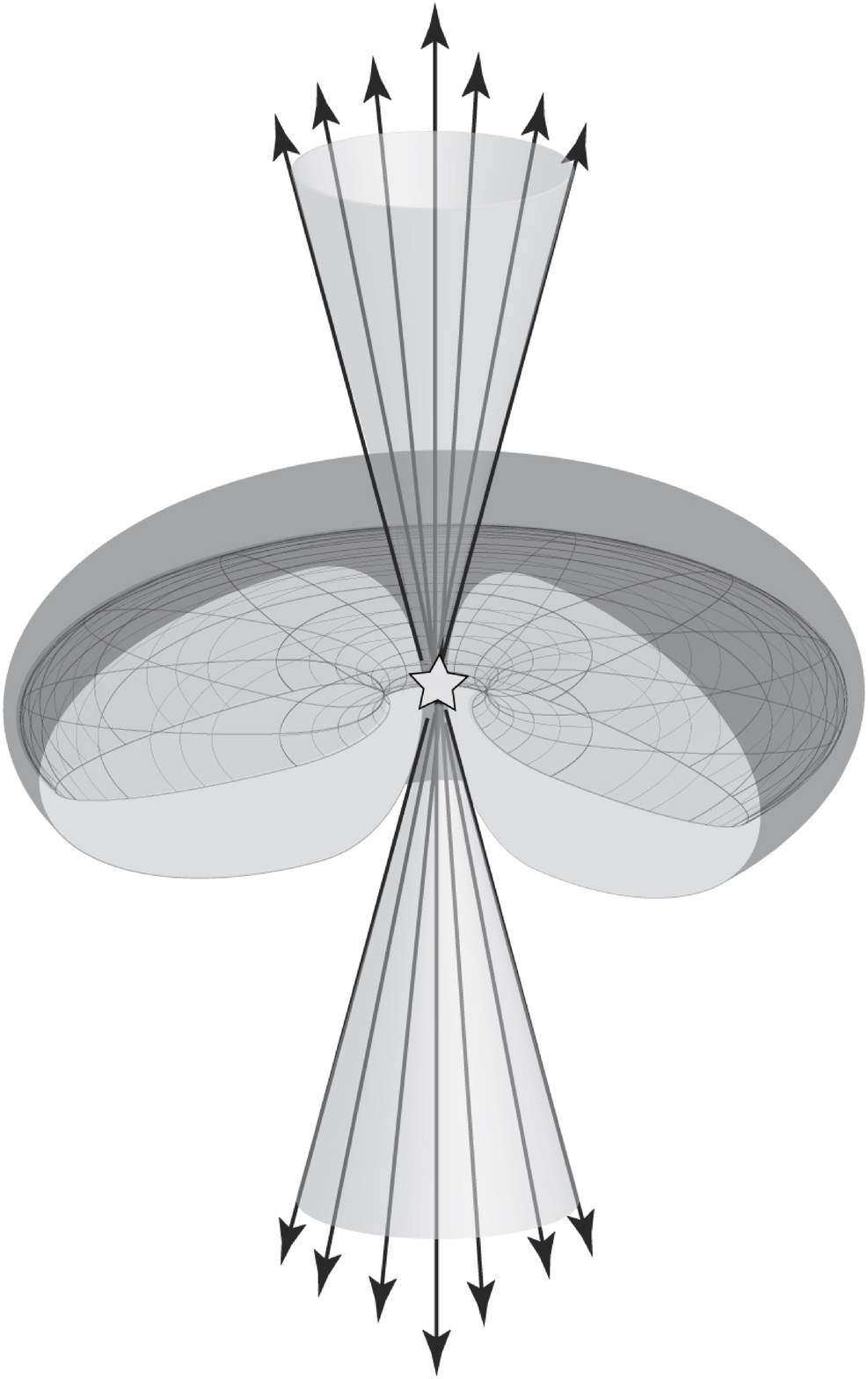}
	\includegraphics[width=0.4\textwidth]{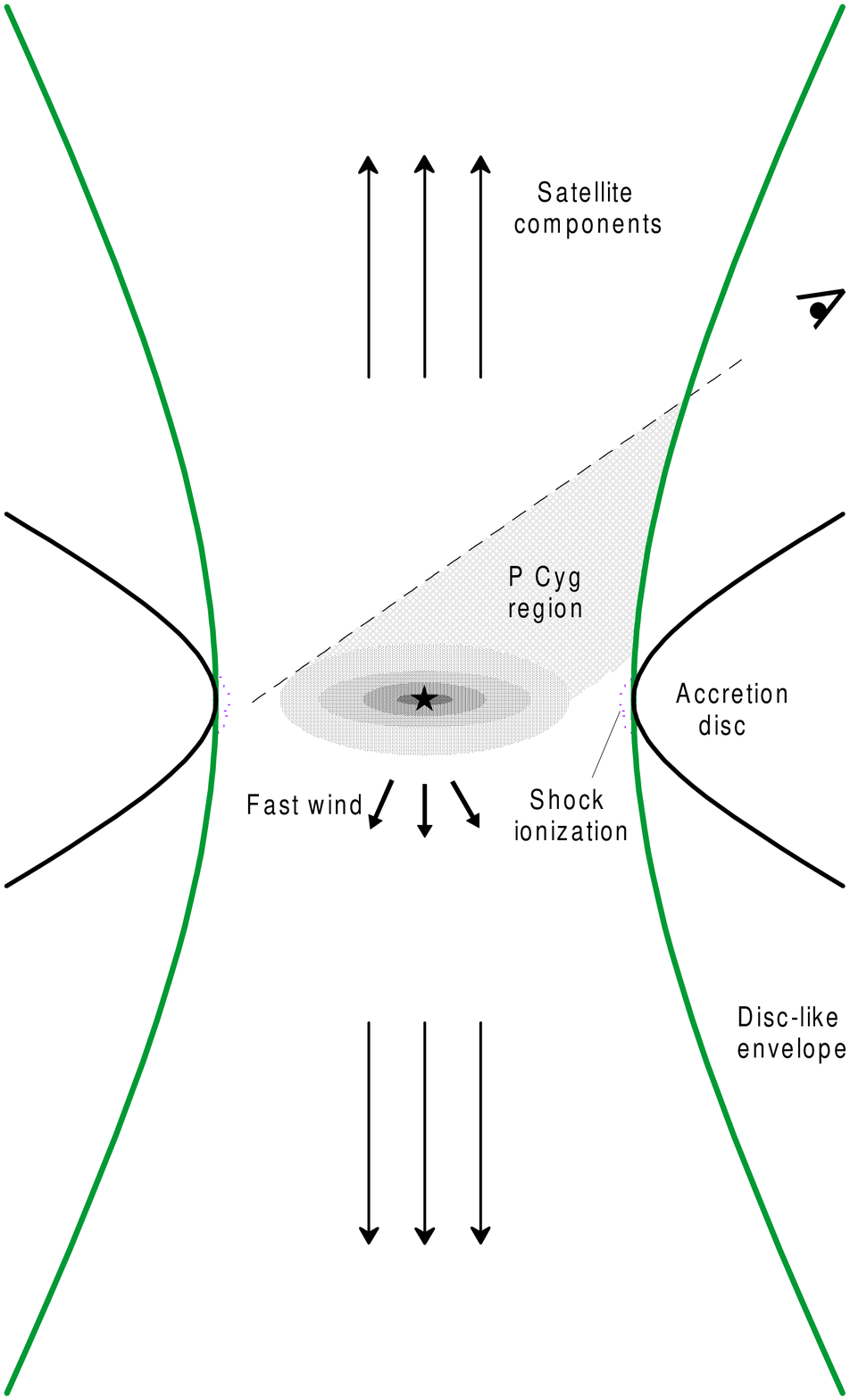}
 \caption{The same like in Fig.~1 but during recurrent strong outburst.}
	\label{model2}
\end{figure}

During the active phase the wind of the compact object ``strips'' the accretion disc and ejects some part of its mass. At the end of each outburst some part of the ejected mass locates in the potential well of the compact object. After the cessation of the wind it begins to accrete again. Because of conservation of the initial angular momentum a disc-like envelope forms surrounding the disc which envelope is located at a greater distance from the orbital plane than the accretion disc itself. The existence of centrifugal barrier leads to the appearance of two hollow cones with a small opening angle ($15\degr$--$30$\degr) around the axis of rotation (see Fig.~\ref{model2}, left panel) \citep{I,BB}.

%%%%%%%%%%%%%%%%%%%%%%%%%%%%%%  FIGURE 2 model2

During the first outburst, disc-like envelope does not exist. The following outbursts during the active phase, the extended disc-like envelope can collimate the wind, which in this case occupies only the two hollow cones and bipolar outflow forms (Fig.~\ref{model2}, left panel). This outflow is observed as high velocity satellite components, situated on either side of the main peak of the emission line. Their presence in the spectrum depends on the density of both the disc-like envelope and the outflowing material. These components will appear only if the density of the disc-like envelope is high enough to provide collimation and the mass-loss rate of the outbursting component is also high. According to our model they are expected to be observed during outbursts accompanied by mass-loss at high rates and preceded by similar strong outburst.

The observed broad emission components indicating an optically thin stellar wind with a high velocity appear close to the compact object where the outflowing material moves in all directions (Fig.~\ref{model2}, right panel). The gas outflowing close to the surface of the cone, whose velocity is lower than that along to the cone axis, can contribute to these emission components too. Depending on the inclination angle of the orbit and the opening angle of the cone the outflow close to the cone surface can have a radial velocity, which makes it able to emit at wavelengths far away from the center of the satellite emission.

\subsection{The 2000--2002 outburst}
\label{subsec:firstout}
%%%%%%%%%%%%%%%%%%%%%%%%%%%%%%  FIGURE 3 Hg_zand
%
\begin{figure}[!htb]
	\includegraphics[width=0.5\textwidth]{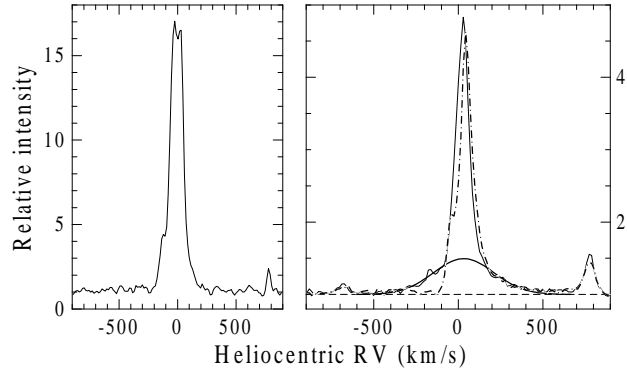}
%    \centering{\epsfig{file=fig3.ps, width=0.5\textwidth}}
 \caption{The profile of the H$\gamma$ line. The spectrum of January 7, 1999 taken in the quiescent state of the system is shown in the left panel.
 					The spectra of November 17 (dot-dash line) and December 5 (solid line), 2000 taken during the outburst phase are shown in the right panel.
 					The Gaussian fit of the broad component is shown too. The level of the local continuum is marked with a dashed line.}	
 \label{Hg_zand}
\end{figure}

We will consider only the H$\gamma$ and \mbox{He\,{\sc i}} profiles of Z~And during its 2000--2002 brightening to illustrate the interpretation of its line spectrum. All of the lines observed by us are considered in the work of \citet{TTB08}.
During this outburst an additional low-intensity emission component of the H$\gamma$ line with a height of $\sim\!\!0.5$ of the local continuum and FWZI of $\sim\!\! 1000$~km\,s$^{-1}$ was seen together with its narrow component (Fig.~\ref{Hg_zand}, right panel). The intensity of the broad emission reached its maximal value at the time of the maximal light whereas the behavior of the narrow component was different. That is why we consider that the H$\gamma$ line consisted in this way of a narrow component of nebular type and a broad component. The broad emission component was analyzed by fitting with Gaussian function.

The blue wing of the broad component was not seen in November 2000 (Fig.~\ref{Hg_zand}) which probably resulted from the absorption by the stellar wind forming the P~Cyg lines (see below). We consider that the broad emission component of the H$\gamma$ line indicates an optically thin stellar wind with a high velocity of $\sim\!\!500$~km\,s$^{-1}$ from the compact object in the system, which wind was also observed in the \mbox{He\,{\sc ii}} $\lambda$\,4686 line.
		
%%%%%%%%%%%%%%%%%%%%%%%%%%%%%%  FIGURE 4 Hei_qa

\begin{figure}[!htb]
	\includegraphics[width=0.5\textwidth]{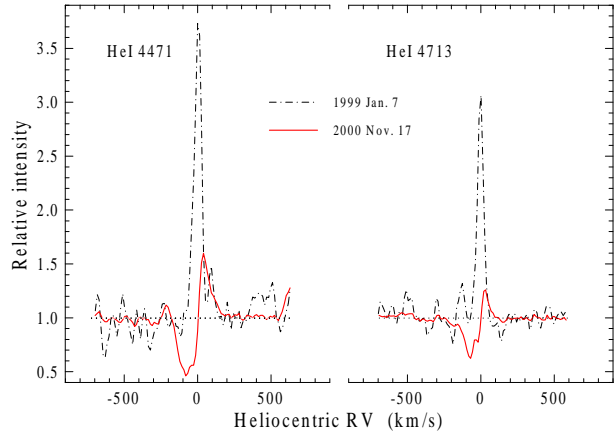}
 \caption{The profiles of the \mbox{He\,{\sc i}} triplet lines in quiescence (January 1999; dot-dashed line) and near the maximum of the first outburst (November 2000; solid line). (From spectroscopy presented in \citet{TTB08}.)}
 \label{Hei_qa}
\end{figure}

\begin{figure}[!htb]
	\includegraphics[width=0.45\linewidth]{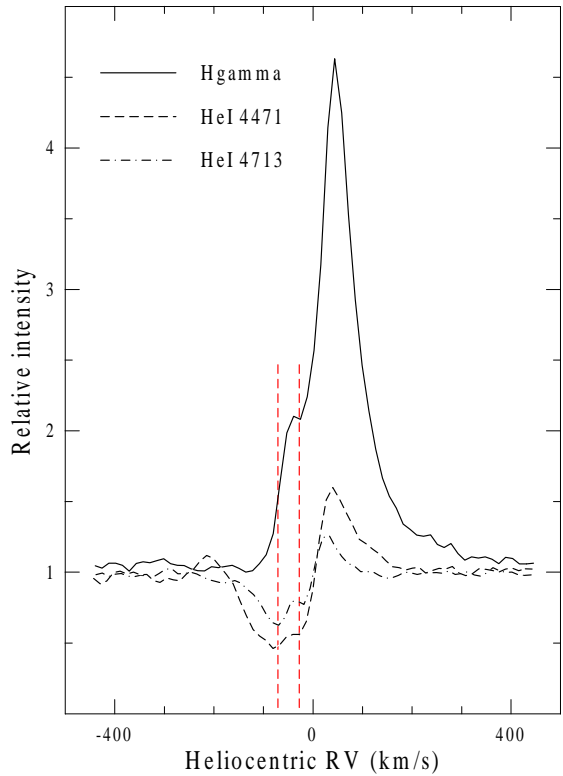}
  \caption{The profiles of the H$\gamma$, \mbox{He\,{\sc i}} $\lambda$\,4471, and \mbox{He\,{\sc i}} $\lambda$\,4713 lines on November 17, 2000.
  The positions of the two components of helium absorption are marked with dashed lines.}
  \label{Hg&Hei}
\end{figure}

The triplet lines of helium \mbox{He\,{\sc i}} $\lambda$\,4471 and \mbox{He\,{\sc i}} $\lambda$\,4713 had purely emission profile and FWHM $= 45$--$50$~km\,s$^{-1}$ in the quiescent state of the system. A blue-shifted P~Cyg absorption with a velocity of $\sim\!\! 60$~km\,s$^{-1}$ appeared in November and December 2000 during the time of the maximal light. In November this absorption had two component structure (Fig.~\ref{Hei_qa}). The \mbox{He\,{\sc i}} $\lambda$\,4471 line reached a residual intensity of $0.46$ in November and $0.60$ in December, respectively. Since the cool giant's continuum at the same time amounted to $\sim\!\!0.07$--$0.08$ of the total continuum of the system at the wavelengths of these lines \citep{TTZ03}, their appearance can be related to optically thick outflow (stellar wind) from the outbursting compact object. Moreover some part of the H$\gamma$ line on November 17, 2000 was probably also absorbed by this wind as far as its absorption dip has the same wavelength position (Fig.~\ref{Hg&Hei}).

%%%%%%%%%%%%%%%%%%%%%%%%%%%%%%  FIGURE 5 Hg&Hei

In the framework of our model the H$\gamma$ and \mbox{He\,{\sc i}} profiles of Z~And can be interpreted in the next way. We suppose that a geometrically thin accretion disc exists in the system in the beginning of the first outburst of the active phase and the inner region of this disc can be optically thick. The high-velocity wind of the compact object arising during the outburst and emitting the broad H$\gamma$ emission component collides with the disc in the region of the orbital plane and its velocity decreases (Fig.~\ref{model1}). At higher stellar latitudes this velocity does not change. This causes optically thick disc-like shell to form, which occults the compact object and plays a role of observed photosphere with an effective temperature much lower \citep{TTT03,Sk06} than that of the compact object in the quiescent state of the system. As a result of the velocity decrease the density of the outflowing gas increases and this part of it which is projected onto the photosphere gives rise to a reversal in the emission profile of H$\gamma$ and to P~Cyg absorption component in the lines of \mbox{He\,{\sc i}}.

We suppose that the line spectrum of any system containing indications of two-velocity regime wind outflow can be interpreted in the framework of the idea for accretion disc surrounding the outbursting compact object and preventing the wind.

\subsection[]{The 2006--2007 outburst}
\label{subsec:fourthout}

\begin{figure}[!tH]%[!htb]%
	\includegraphics[width=0.5\linewidth]{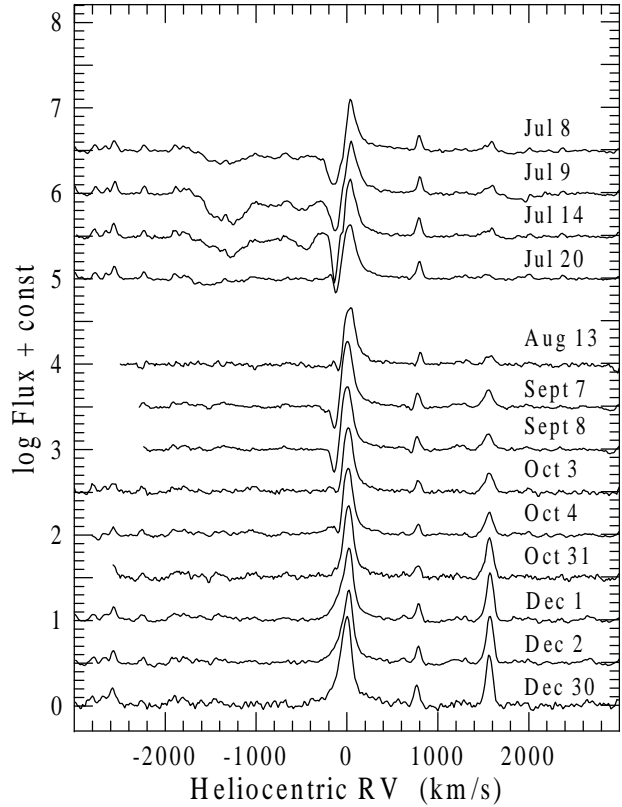}
\caption{Time evolution of the H$\gamma$ line of Z~And.}
  \label{Hg_2006}
\end{figure}

\begin{figure}[!tH]
	\includegraphics[width=0.5\linewidth]{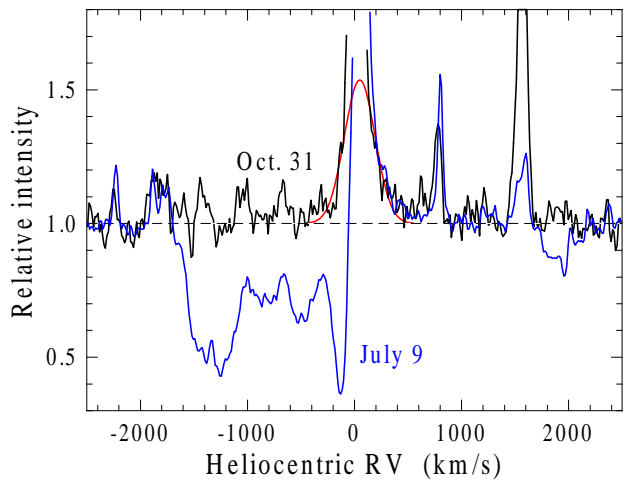}
\caption{The profile of the H$\gamma$ line on July 9 and October 31. The Gaussian fit of the broad component is also shown. The level of the local continuum is marked with a dashed line.}
  \label{Hg_2006_2}
\end{figure}

During the major 2006 outburst the H$\gamma$ profile of Z~And was multicomponent one indicating all regimes of the outflow in the system and to illustrate the interpretation of its line spectrum we will use only the H$\gamma$ line. The H$\gamma$ line presented a broad emission component with low intensity and FWZI of $\sim\!\!1000$~km\,s$^{-1}$ in addition to its central narrow component with a nebular profile. The broad component is best seen on the spectra taken after October 31, 2006 (see Figs.~\ref{Hg_2006} and ~\ref{Hg_2006_2}). The data obtained in this period of time show that the energy flux of the broad component decreased when the light weakened after its maximum, whereas the behavior of the central narrow emission was different \citep{TTB12}. The broad component was analyzed by fitting with a Gaussian function (Fig.~\ref{Hg_2006_2}) \citep{TTB12}. On the spectra taken in
July--September the blue wing of the broad component was not seen because of blending with the P~Cyg absorption component (see below and Fig.~\ref{Hg_2006}). On the spectra taken in October and December the blue wing appeared to be less extended than the red one due to blending with the P~Cyg absorption. We consider that the broad emission component indicates an optically thin stellar wind with a high velocity of $\sim\!\!500$~km\,s$^{-1}$ from the compact object in the system.

%%%%%%%%%%%%%%%%%%%%%%%%%%%%%%  FIGURE 6 Hg_2006
%

On the spectra taken during July--September 2006 the central narrow component of the line had positive radial velocity, which was due to the presence of a blueshifted P~Cyg absorption (Figs.~\ref{Hg_2006} and ~\ref{Hg_2006_2}). In July this absorption presented multi-component structure and occupied a velocity range from $\sim\!\!\!100$ to $1\,500$--$1\,600$~km\,s$^{-1}$. After that it gradually weakened and converted in low-velocity absorption presenting in the spectrum until the beginning of October 2006 (Fig.~\ref{Hg_2006}). The residual intensity of this absorption was minimal in the middle of July at $0.4$. As the cool giant's continuum, at the same time, was less than $9$ per cent of the total continuum of the system at the wavelength position of the $B$ photometric band \citep{Sk09} which is close to the H$\gamma$ line, the P~Cyg absorption may be related to the optically thick mass outflow from the outbursting compact object.

The comparison of the spectra taken on July 9 and October 31 shows that the red wing of the broad component on the two spectra coincide (Fig.~\ref{Hg_2006_2}), which suggests that the H$\gamma$ line has three components, consisting of a central narrow emission component, a broad emission component with low-intensity and multi-component P~Cyg absorption occupying broad region of velocities of the outflowing material---from $\sim\!\!\!100$ to $1\,500$--$1\,600$~km\,s$^{-1}$.

%%%%%%%%%%%%%%%%%%%%%%%%%%%%%%  FIGURE 7 Hg_2006_2

\begin{figure}[!htb]%[!tH]
	\includegraphics[width=0.5\linewidth]{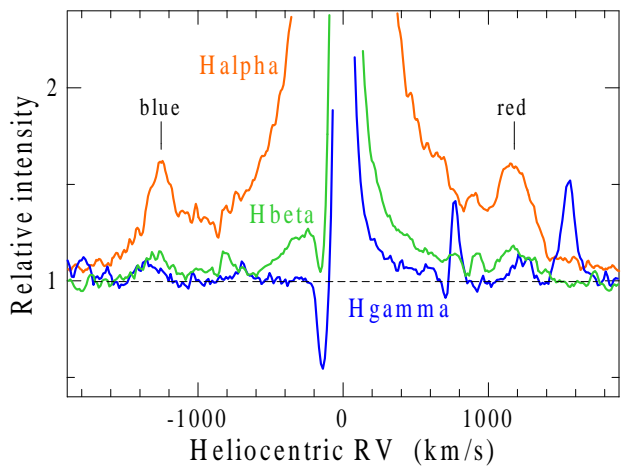}
%	\vspace{4cm}
\caption{The profiles of the H$\alpha$, H$\beta$, and H$\gamma$ lines of Z~And based on a CCD frame on September 8, 2006. The satellite components are pointed with vertical lines. The level of the local continuum is marked with a dashed line.}
  \label{Habg_8sep}
\end{figure}

The comparison of the H$\gamma$ profile with those of H$\alpha$ and H$\beta$ taken on September 8 (Fig.~\ref{Habg_8sep}) shows that the H$\gamma$ line has satellite emission components, too, with velocities very close to the H$\alpha$ and H$\beta$ velocities. Both H$\gamma$ satellite components were observed only in September when the intensity of the H$\alpha$ components was maximal. In October only a blueshifted H$\gamma$ emission presented in the spectrum. This means that in September the H$\gamma$ profile consisted of four groups of components.

The H$\gamma$ profile is considered in the light of the second variant of the model (Fig.~\ref{model2}) where a disc-like envelope surrounding the accretion disc exists in the system. The four-component line can be interpreted in the following way. The high velocity wind indicated by the broad emission component collides with the disc and disc-like envelope and a collimated outflow forms after the collision. The outflowing gas which is projected onto the observed photosphere of the outbursting compact object (the disc-like shell) gives rise to the P~Cyg absorption (Fig.~\ref{model2}). Due to this absorption the blue wing of the broad emission component is not seen. The radial velocities in the area of the wind projecting onto the observed photosphere cover an appreciable range---from values close to zero to the maximal observed velocity of the collimated outflow. This provides possibility broad absorption to form. The red wing of the broad emission is seen since some part of the back wind component is not occulted by the observed photosphere (the disc-like shell). The high-velocity satellite components arise in more outer regions of the bipolar outflow.

%%%%%%%%%%%%%%%%%%%%%%%%%%%%%%  FIGURE 8 Habg_8sep

We suppose that the line spectrum of any system containing indications of both stellar wind and collimated outflow can be interpreted in the framework of the idea for collimated stellar wind.

\section{The spectral variability of other classical symbiotic stars}
\label{sec:otherstars}

\subsection{HEN~3-1341}
\label{subsec:hen3}
	
The Hen~3-1341 system consists of a cool giant of spectral type M4 \citep{MS} without circumstellar dust \citep{Munari+92,tomata00}, very hot and luminous white dwarf with effective temperature of $\sim\!\!\!1.2 \times 10^5$~K and luminosity of $3.8 \times 10^3$~L$_{\sun}$, and an surrounding nebula partly photoionized by the white dwarf \citep{Munari+05}.

%%%%%%%%%%%%%%%%%%%%%%%%%%%%%%  FIGURE 9 hen3_lc

\begin{figure}[!htb]%[!tH]
	\includegraphics[width=0.65\linewidth]{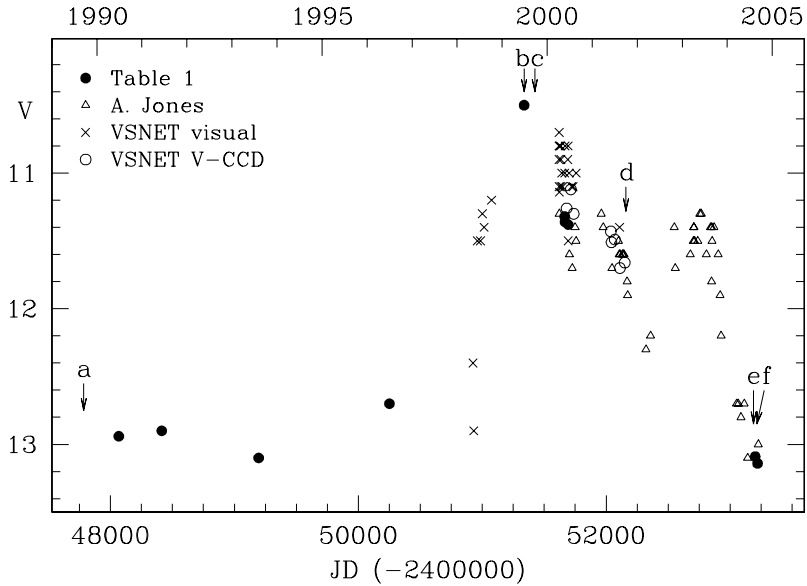}
\caption{The V light curve of Hen~3-1341 over the time 1990--2005 showing the 1998--2004 outburst from the paper of \citet{Munari+05}.}
  \label{hen3_lc}
\end{figure}

%%%%%%%%%%%%%%%%%%%%%%%%%%%%%%  FIGURE 10

\begin{figure}[!tH]
	\includegraphics[width=.8\linewidth]{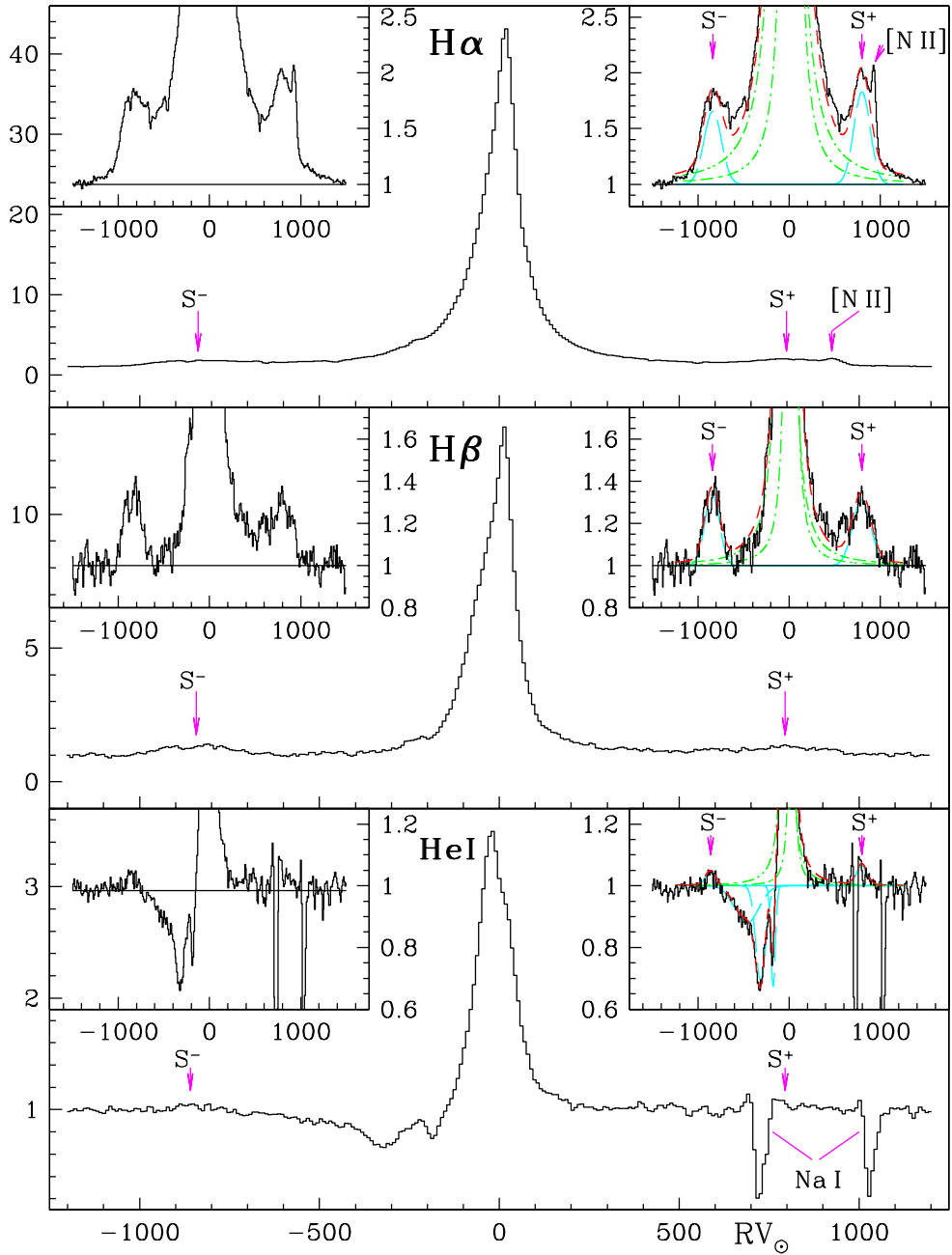}
\caption{The profiles of the H$\alpha$, H$\beta$, and \mbox{He\,{\sc i}} $\lambda$\,5876 lines of Hen~3-1341 based on CCD frames on June 8, 1999 (from the article of \citet{tomata00}).}
  \label{hen3_tomata}
\end{figure}

The system underwent a large outburst lasting from 1998 to 2004 (Fig.~\ref{hen3_lc}) which was its first outburst ever recorded \citep{tomata00,Munari+05}. High-resolution spectral data were obtained by \citet{tomata00} and \citet{Munari+05} during this outburst. The data of \citet{tomata00} taken on June 8, 1999 shows that the H$\alpha$ line had an emission profile consisting of an intensive central singlepeaked component and additional satellite components with velocity of about 800~km\,s$^{-1}$ on both sides of the central emission. The same appearance had the H$\beta$ profile whose satellite components had the same velocity. The profile of the triplet \mbox{He\,{\sc i}} $\lambda$\,5876 line was similar to those of the Balmer lines consisting of the same components but this line contained two-component P~Cyg absorption in addition which occupied broad velocity range---from about $150$ to about $700$~km\,s$^{-1}$ (Fig.~\ref{hen3_tomata}). The high-velocity emission satellite components of all lines were attributed by the authors to bipolar jets and the \mbox{He\,{\sc i}} P~Cyg absorption was recognized as signature of mass outflow (stellar wind) from the outbursting compact object. \citet{Munari+05} paid attention to the tight correlation between the strength of the H$\alpha$ satellite components and the \mbox{He\,{\sc i}} P~Cyg absorption. They came to the conclusion that the wind plays a role of feeding mechanism for the bipolar jets.

In our opinion the correlation between the H$\alpha$ satellite emissions and the \mbox{He\,{\sc i}} absorption of Hen~3-1341 is of primary importance. According to the modern theory, the existence of bipolar collimated jets is supposed to be due to the presence of  magnetic disc which transforms the potential energy of the accreting material into kinetic energy of the outflowing gas. This means that the accreting material provides the jet outflow. If we observe simultaneous indications of bipolar outflow and stellar wind, accretion and stellar wind from the white dwarf should happen at the same time. To avoid this difficulty we use another model for the interpretation of the spectrum of Hen~3-1341, namely one with collimated stellar wind.

In the framework of this model the \mbox{He\,{\sc i}} $\lambda$\, 5876 line can be interpreted in the following way. The stellar wind from the white dwarf collides with the disc and disc-like envelope and a bipolar collimated outflow forms after the collision. The area of the wind which is projected onto the effective photosphere of the outbursting compact object gives rise to the P~Cyg absorption. The radial velocities in this area cover an appreciate range---from values close to zero to the real wind velocity (Fig.~\ref{model2}). This provides the possibility a broad absorption to form. The high-velocity satellite emissions of all lines appear in more outer regions of the wind, not projecting onto the effective photosphere.

\begin{figure}[!tH]
	\includegraphics[width=.7\textwidth]{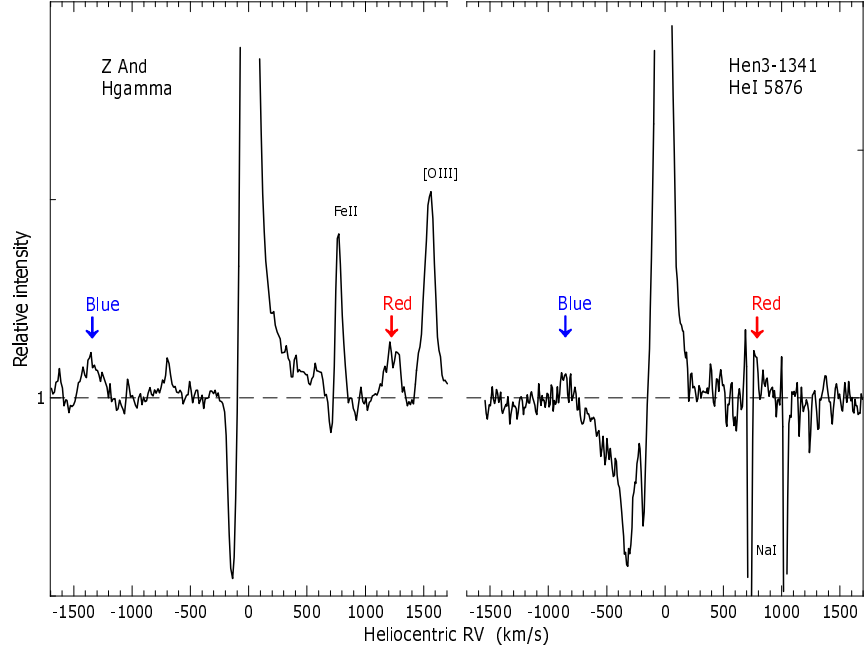}
\caption{The H$\gamma$ profile of Z~And based on a CCD frame on September 8, 2006 and the \mbox{He\,{\sc i}} $\lambda$\,5876 profile of Hen 3-1341 based on CCD frame on June 8, 1999 \citep{tomata00}.}
  \label{hen_zand}
\end{figure}

The \mbox{He\,{\sc i}} profile of Hen~3-1341 is compared with the H$\gamma$ profile of Z~And in Fig.~\ref{hen_zand}. It is seen the great identity of these profiles which gives us a reason to interpret them in the framework of the same model.

%%%%%%%%%%%%%%%%%%%%%%%%%%%%%%  FIGURE 11 hen_zand

\subsection{BF~Cyg}
\label{subsec:bfcyg}

The symbiotic BF~Cyg system is an eclipsing binary \citep{Bel00} consisting of a late-type cool component classified as an M5\,III giant \citep{KFC87}, a hot compact object with temperature of about $100\,000$~K and an extended surrounding nebula \citep{Sk05}. Its orbital period is about $757^{\rm d}$, which is based on both photometric \citep{Mik87} and radial velocity \citep{Fekel01} data.

\begin{figure*}[!htb]%[!tH]
	\includegraphics[width=0.38\textwidth,angle=-90]{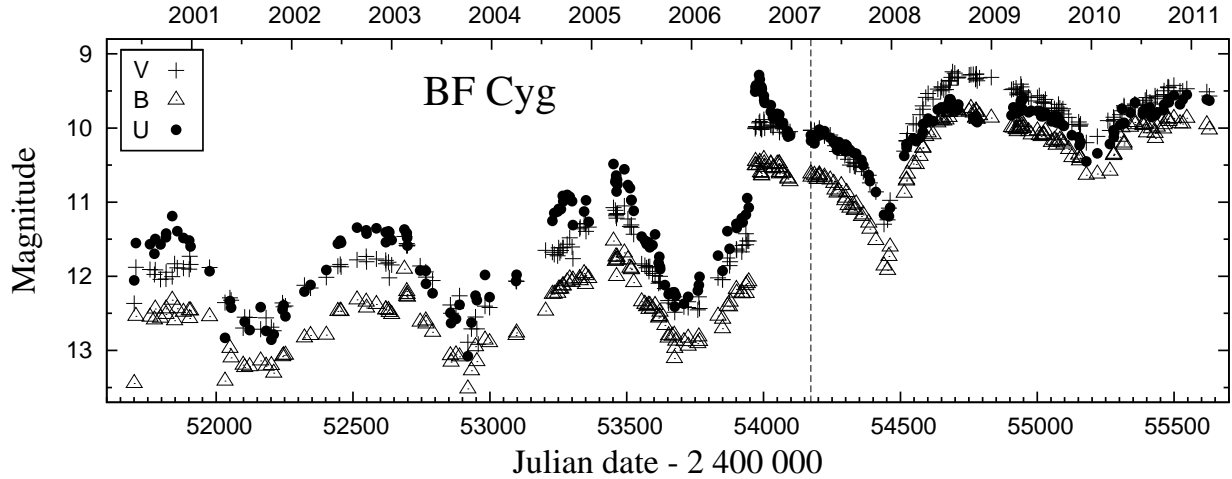}
%	\vspace{4cm}
\caption{The UBV light curves of BF~Cyg during its last outburst \citep{Sk12} (Skopal et~al.\ 2012).}
  \label{bfcyg_lc}
\end{figure*}

The historical light curve of BF~Cyg shows three types of activity. It contains one very prolonged outburst, lasting for decades and similar to that of  the symbiotic novae, several eruptions of an other type such as those observed in the classical symbiotic stars and sudden rapid brightenings, lasting a small portion of the orbital period \citep{Sk97}. The last major eruption of BF~Cyg began in 2006 and at the present time continues (Fig.~\ref{bfcyg_lc}).

%%%%%%%%%%%%%%%%%%%%%%%%%%%%%%  FIGURE 12 bfcyg_lc

\begin{figure}[!tH]
	\includegraphics[width=.48\linewidth]{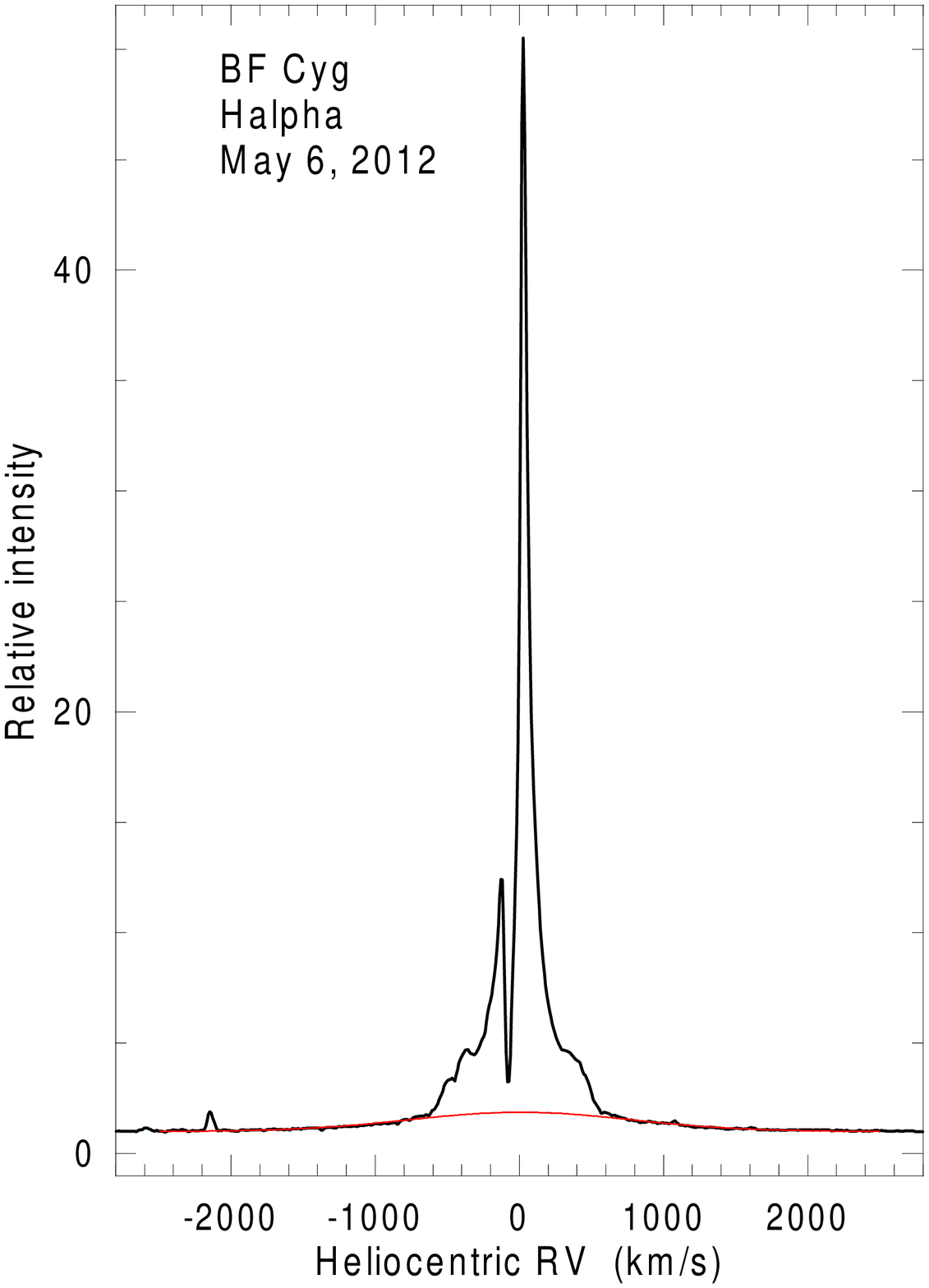}
	\includegraphics[width=.468\linewidth]{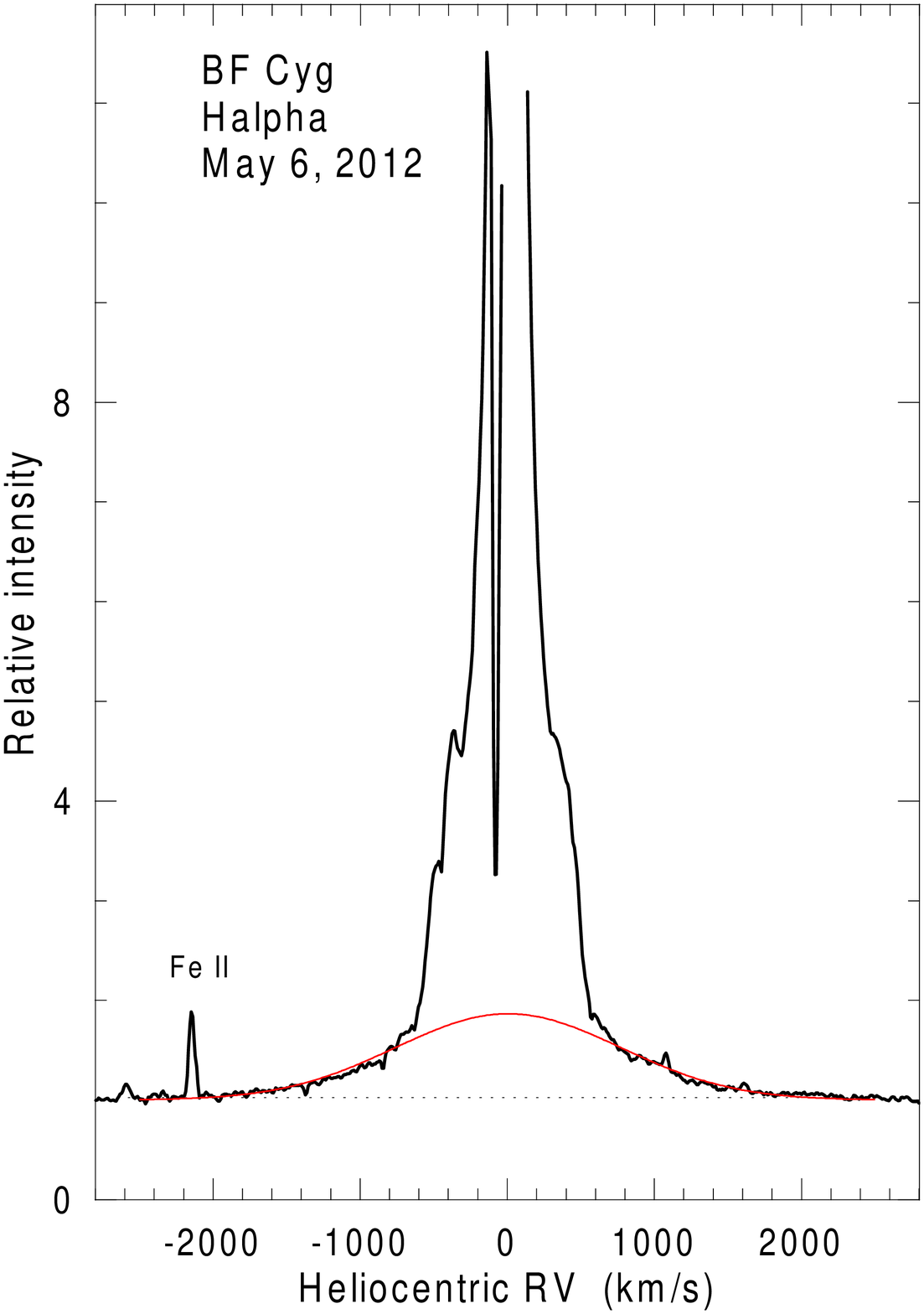}
\caption{\emph{Left panel}: The profile of the H$\alpha$ line of BF~Cyg based on a CCD frame on May 6, 2012. \emph{Right panel}: The area of the wings where the broad component is better seen. The level of the local continuum is marked with a dashed line.}
  \label{bfcyg_ha}
\end{figure}

\begin{figure}[!tH]
	\includegraphics[width=.47\textwidth]{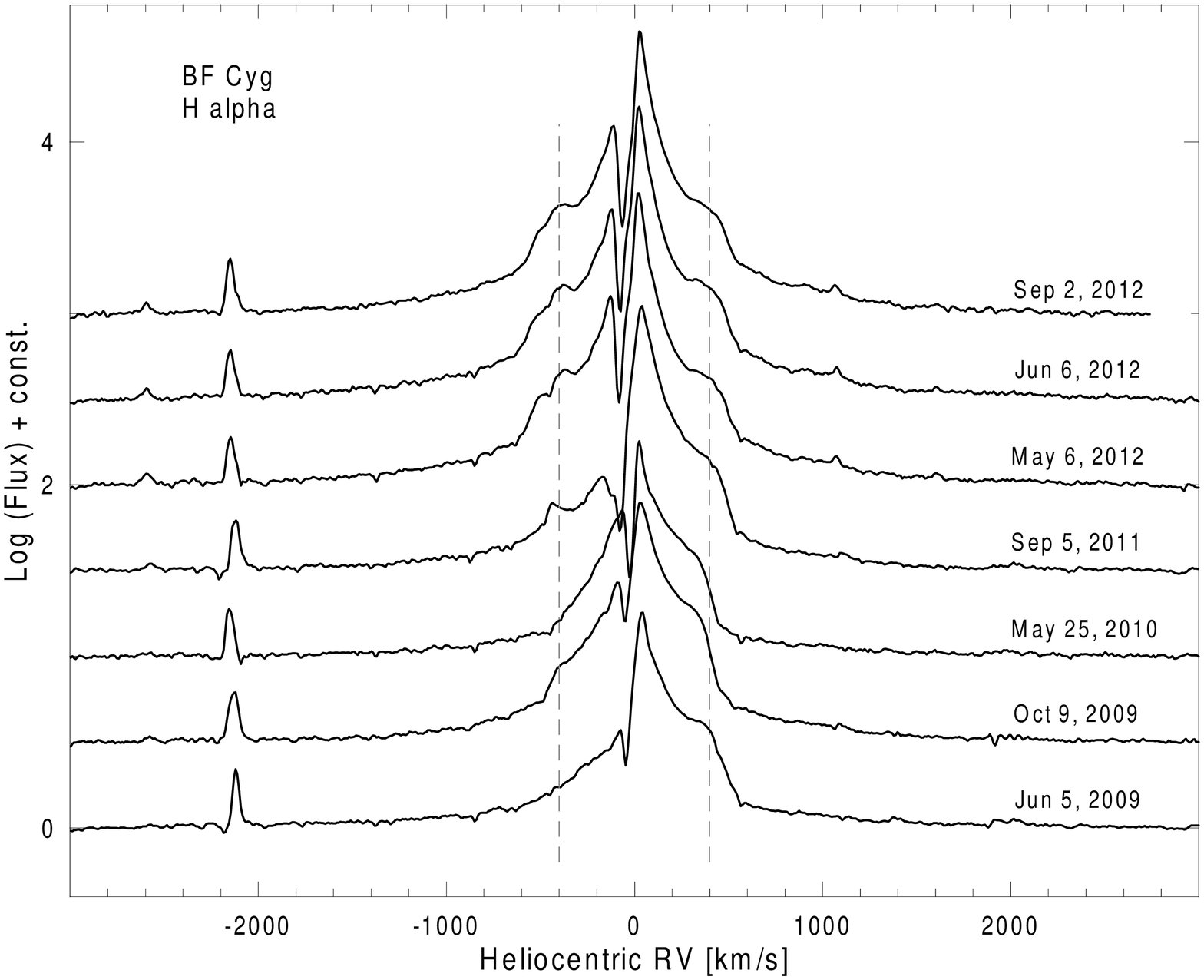}
	\hspace{.4cm}
	\includegraphics[width=.47\textwidth]{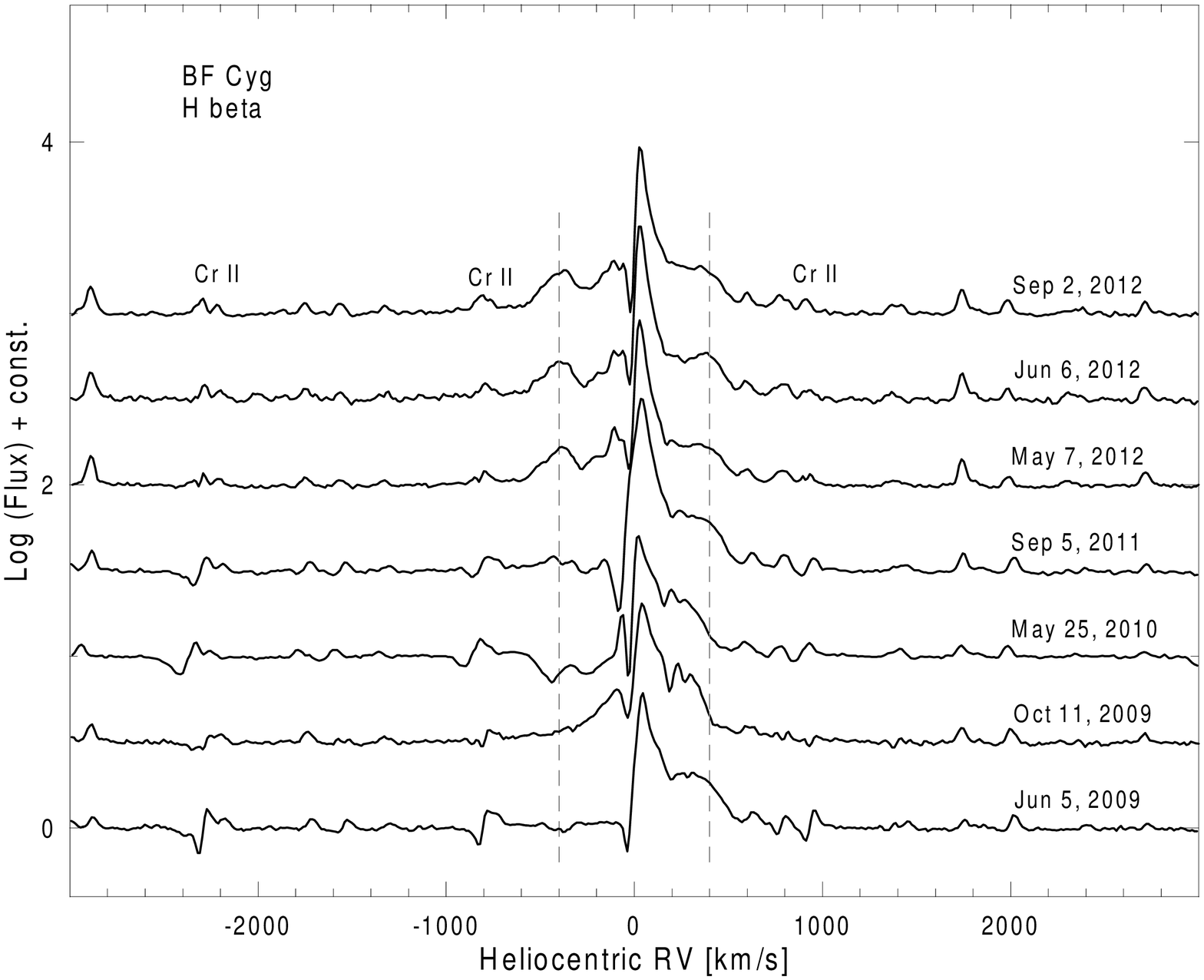}
\caption{Time evolution of the H$\alpha$ and H$\beta$ lines of BF~Cyg.}
  \label{bfcyg_hab_evol}
\end{figure}

We observed the regions of the H$\alpha$ and H$\beta$ lines of the spectrum of BF~Cyg on seven nights during 2009 June--2012 September. During the whole time of our observations the H$\alpha$ line of the BF~Cyg system had multicomponent profile. It had central narrow emission component which at times was double with a blueshifted central reversal and very broad wings with a low intensity extended to velocity of about $\pm\,2200$~km\,s$^{-1}$ from the center of the line whose nature is rather unclear. These wings were analyzed by fitting with a Gaussian function.  The H$\alpha$ line had an additional emission, which after May 6, 2012 contained two peaks, situated on either side of its central narrow component (see Fig.~\ref{bfcyg_ha}). The comparison with the H$\beta$ line shows that these peaks had the same wavelength position of $\sim\!\!400$~km\,s$^{-1}$ like the satellite components of H$\beta$ indicating bipolar collimated outflow (see below). That is why we suppose that this outflow gave rise to the H$\alpha$ peaks too (Fig.~\ref{bfcyg_hab_evol}).

%%%%%%%%%%%%%%%%%%%%%%%%%%%%%%  FIGURE 13 bfcyg_ha

%%%%%%%%%%%%%%%%%%%%%%%%%%%%%%  FIGURE 14 bfcyg_hab_evol

The H$\beta$ line consisted of a narrow core, which at times was double with a blueshifted central reversal similar to H$\alpha$, and broader multipeaked emission (Fig.~\ref{bfcyg_hab_evol}).

The high-velocity absorption component on the spectrum of May 25, 2010 and the most blueshifted emission peak on the last three spectra have the same position close to $-400$~km\,s$^{-1}$. This gives us a reason to suppose they are related to the same individual line component which appears as absorption at some stage of the outburst and after that goes into emission. The transformation of the absorption into emission is seen on the spectrum of Sept 5, 2011. The presence of this individual component in the spectrum of Jun 5, 2009 is indicated by one weak absorption. This absorption does not present in the spectrum of Oct 11, 2009 taken at orbital phase $0.914$, probably because of eclipse effect of the effective photosphere (pseudophotosphere) of the expanding compact object.

We assume the individual component at a position of $-400$~km\,s$^{-1}$ probably indicates collimated ejection from the expanding compact object which ejection is optically thick at the more early stage of the outburst. On the spectra taken after May 2012 one redshifted emission peak at the same velocity of $400$~km\,s$^{-1}$ is seen. Then we assume that it is emitted by the back component of the collimated ejection, i.e. the ejection is bipolar one. This bipolar ejection probably gives rise to the peaks/shoulders at the same velocity of $\sim\!\!400$~km\,s$^{-1}$ on the other H$\beta$ spectra and to the peaks/shoulders in the emission profile of the H$\alpha$ line. The satellite H$\alpha$ and H$\beta$ emission components with a velocity of about $\pm 400$~km\,s$^{-1}$ are analyzed with a Gaussian function to determine their exact wavelength position and equivalent width in the paper by \citet{Sk_prep}. During the 2006--2012 outburst of BF~Cyg satellite line components indicating collimated ejection were observed for the first time in this system.

One absorption component at a velocity position of $-250$~km\,s$^{-1}$ indicating P~Cyg wind is seen on the spectrum of May 25, 2010. After that time it weakened and was not present in the spectrum of May 7, 2012. Another P~Cyg absorption component with lower velocity of $50$--$100$~km\,s$^{-1}$ was seen in the spectra till Sept 5, 2011 and after that disappeared too. This indicates fading of the P~Cyg wind together with the decrease of the optical thickness of the collimated outflow.

\begin{figure}[!htb]%[!tH]
	\includegraphics[width=.7\textwidth]{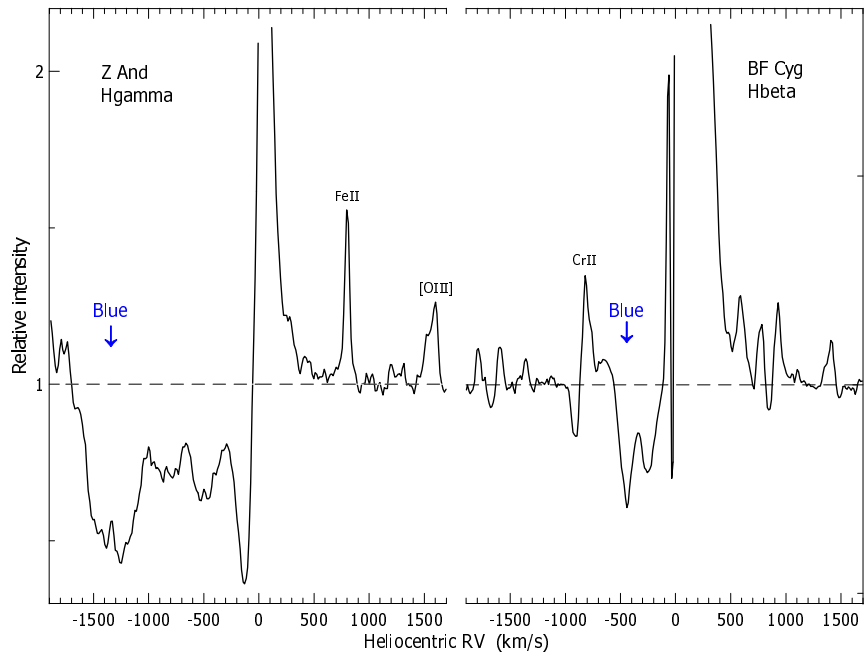}
\caption{The H$\gamma$ profile of Z~And based on a CCD frame on July 9, 2006 and the H$\beta$ profile of BF~Cyg based on a CCD frame on May 25, 2010.}
  \label{bfcyg_zand}
\end{figure}

The H$\beta$ profile of BF~Cyg is compared with the H$\gamma$ profile of Z~And in Fig.~\ref{bfcyg_zand}. Both systems have an absorption satellite component and additional P~Cyg absorption.

%%%%%%%%%%%%%%%%%%%%%%%%%%%%%%  FIGURE 15 bfcyg_zand

The H$\beta$ profile of BF~Cyg is considered in the light of the same scenario which was proposed for interpretation of the Z~And and Hen~3-1341 lines.

\subsection{AG DRA}
\label{subsec:agdra}

The AG~Dra star is a known symbiotic system with a low metal abundance, large barycentric velocity ($\gamma = -148$~km\,s$^{-1}$) and high galactic latitude ($b=41$\degr) belonging thus to the old halo population \citep{Smith96}. It belongs to the group of yellow symbiotics and consists of a cool primary of the spectral type K, probably more luminous than a normal class III giant [31--33], a hot compact object with a high luminosity and temperature of $1.0 \times 10^{-5}$--$1.7 \times 10^{-5}$~K \citep{Mik95,Gr97} and an ionized circumbinary nebula. The photometric orbital period of the system is about $550^{\rm d}$ \citep{Meinunger79,Sk94}, confirmed by measurements of the radial velocity variations of the cool component \citep{Mik95,Smith96}.

\begin{figure*}[!htb]
	\includegraphics[width=0.38\textwidth,angle=-90]{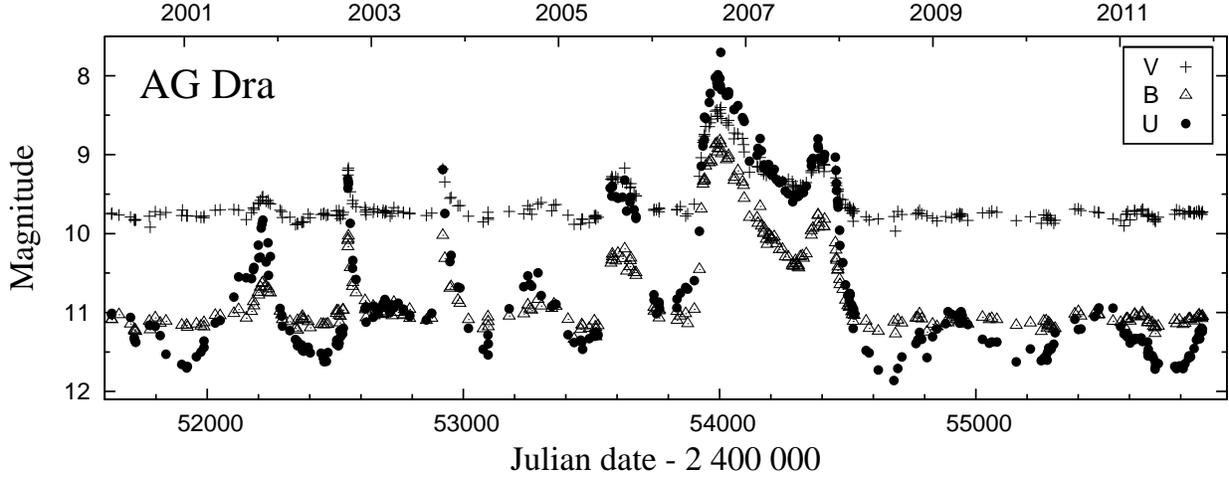}
\caption{The UBV light curves of AG~Dra during its last active phase \citep{Sk12}.}
  \label{agdra_lc}
\end{figure*}

AG~Dra belongs to the group of symbiotics that show a ``classical'' or Z~And type outburst. It has undergone several active phases (after 1936, 1951, 1966, 1980, and 1994), characterized by one or more light maxima. The last active phase of AG~Dra began in 2001 and prolonged till 2008 consisting of several optical outbursts (Fig.~\ref{agdra_lc}).

%%%%%%%%%%%%%%%%%%%%%%%%%%%%%%  FIGURE 16 AGDra agdra_lc

%%%%%%%%%%%%%%%%%%%%%%%%%%%%%%  FIGURE 17 AGDra agdra_heii

\begin{figure*}%[!tH]
%	\vspace{4cm}
	\includegraphics[width=0.7\textwidth]{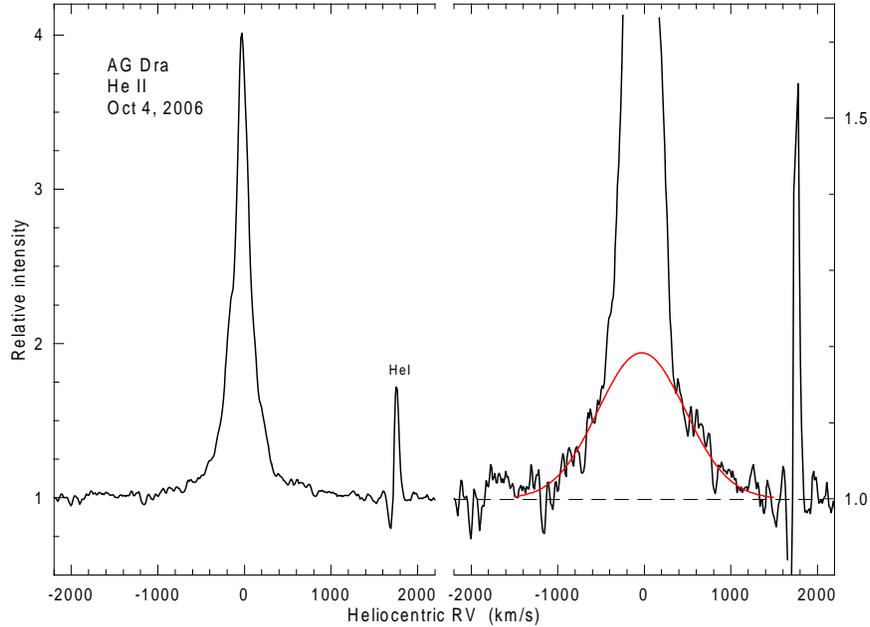}
\caption{\emph{Left panel}: The profiles of the \mbox{He\,{\sc ii}} $\lambda$\,4686 and \mbox{He\,{\sc i}} 						 $\lambda$\,4713 lines of AG~Dra based on a CCD frame on October 4, 2006. \emph{Right panel}: The area of the wings of the \mbox{He\,{\sc ii}} $\lambda$\,4686 line where its broad component is seen. The level of the local continuum is marked with a dashed line.}
\label{agdra_heii}
\end{figure*}

We observed the system at the time of the light maximum of the strongest of the outbursts---the major 2006 one, taking spectrum in the region of the \mbox{He\,{\sc ii}} $\lambda$\,4686 line on October 4, 2006 (JD~2\,454\,013.25). At the time of our observation the triplet line of helium, \mbox{He\,{\sc i}} $\lambda$\,4713, of AG~Dra had a blue-shifted absorption of the type P Cyg with observed velocity of $-230$~km\,s$^{-1}$ (Fig.~\ref{agdra_heii}, left panel). After subtraction of the system's velocity of $-148$~km\,s$^{-1}$ the velocity of this absorption amounted to about $-80$~km\,s$^{-1}$. The depth of this absorption was $0.22$ of the continuum of the system. The continuum of the outbursting hot component at that time amounted to $0.02$--$0.03$ of the total continuum of the system at the wavelength position of the line and the giant's continuum---to about $0.35$--$0.40$ (Skopal, private communication). Then the appearance of the P~Cyg absorption component of the \mbox{He\,{\sc i}} $\lambda$\, 4713 line can be related to the wind of the giant.

At the time of our observation the \mbox{He\,{\sc ii}} $\lambda$\,4686 line of AG~Dra consisted of two emission components---a central narrow component and a broad component with low intensity whose width at the level of the local continuum (FWZI) reached $2520$~km\,s$^{-1}$. The broad component was analyzed by fitting with a Gaussian function (Fig.~\ref{agdra_heii}, right panel) as described in the work of \citet{TT02}. The error of the equivalent width obtained with this procedure is $7$ per cent and it depends primarily on the error of the local continuum. The velocity of the stellar wind (see below) of the hot outbursting component of the system is equal to the half of the FWZI of the line.

A weak broad emission component of the \mbox{He\,{\sc ii}} $\lambda$\,4686 line of AG~Dra exists at some times in quiescent state \citep{GR99} and is determined most probably by electron scattering \citep{TT02}. We should treat different mechanisms of line broadening to conclude about the nature of the broad component of this line during the 2006 outburst. First we will consider the possibility that it is determined only by electron scattering. The absolute flux of the line was calculated using its equivalent width and the continuum flux at its position. The continuum flux was obtained using linear interpolation of the B and V photometric fluxes taken on close nights from the paper of \citet{Sk07}. The uncertainty of the continuum flux is not more than $10$ per cent. The B and V fluxes were corrected for the interstellar reddening of $E(\mathrm{B}-\mathrm{V})=0.06$ \citep{Mik95,Gr97,GR99}. The total flux of the line, which is a sum of the fluxes of the two components, is $2.782\times10^{-12}$ erg\,cm$^{-2}$\,s$^{-1}$, which gives emission measure of $1.69 \times 10^{59}$ (d/1.7~kpc)$^{2}$ cm$^{-3}$. This emission measure was calculated with using a recombination coefficient for case B, corresponding to electron temperature of $30\,000$~K and a number density of $10^{12}$~cm$^{-3}$ \citep{SH} since according to \citep{Sk+09} the electron temperature and the mean density in the close vicinity of the compact object during outburst are about $29\,000$~K and $10^{12}$~cm$^{-3}$, respectively. Accepting such a  density, we obtain a radius of 3.43 $\times$ 10$^{11}$ (d/1.7~kpc)$^{2/3}$ cm of the spherical emitting volume. If the broad component was produced only by electron scattering it would be risen in a region with an optical thickness of $0.30$. Using this value and a number density of $10^{12}$~cm$^{-3}$, we derive a radius of $4.57 \times 10^{11}$(d/1.7~kpc)$^{2/3}$~cm, which is very close to that of $3.43 \times 10^{11}$(d/1.7~kpc)$^{2/3}$~cm based on the observed emission measure. This means that the broad component could be due only to electron scattering.

However, we should consider other mechanisms of line broadening too since the broad component is possible to be formed by more than one of them.  An emission line similar to the \mbox{He\,{\sc ii}} $\lambda$\,4686 broad component can be radiated by a small accretion disc with low luminosity appeared from wind accretion and rotating around the compact object with Keplerian velocity. To consider this possibility we need the size of the outbursting compact component of AG~Dra. According to the results of the analysis of the continuum energy distribution of the AG~Dra system at the time of its 2006 light maximum (Skopal, private communication) the outbursting component maintained the same features of a compact object with a radius of $0.10\,R_{\sun}$ like during the 1994 light maximum. This radius, however, is for a distance of $1.1$~kpc. We use a distance of $1.7$~kpc according to the paper by \citet{TTI00} and the radius in this case should be increased to $0.15\,R_{\sun}$. Then for masses of $0.3$--$0.6\, M_{\sun}$ \citep{Mik95,GR99} a Keplerian velocity of $\sim\!\!620$--$870$~km\,s$^{-1}$ is obtained. This velocity is smaller than the velocity of $1260$~km\,s$^{-1}$ based on the HWZI of the broad component. Since, however, the distance to the system is obtained with some uncertainty and it could be smaller than $1.7$~kpc, the radius on its side could be smaller than $0.15\, R_{\sun}$ and the Keplerian velocity---greater than calculated one. Then our supposition for a disc origin of this component can not be thus rejected.

Another possibility is the broad component of the \mbox{He\,{\sc ii}} $\lambda$\,4686 line to be emitted by a high-velocity stellar wind. According to \citet{GR99} the hot compact object in the AG~Dra system expands during active phases. Then we suppose that the broad component can be emitted by an optically thin stellar wind too. In this case we should accept all three possibilities---notably that a scattering by free electrons as well as emission of Keplerian accretion disc and optically thin stellar wind give rise to a broad component. Then the wind should collide with the disc which can make possible a disc-like formation surrounding the compact object to appear in the system as in the case of Z~And. This disc-like formation, however, could be optically thin and difficult to be detected observationally.

\section[]{Conclusions}
\label{sec:concl}

We consider profiles of selected lines of several symbiotic systems which contain high-velocity satellite components indicating bipolar collimated outflow from the outbursting compact object in these systems along with blueshifted absorption indicating P~Cyg wind from the same object. We interpret these profiles in the framework of the scenario proposed for explanation of the line profiles of Z~And during its 2000--2012 active phase \citep{T+11,TTB12}.

The profile of the H$\gamma$ line of Z~And during its 2006 outburst consisted of four groups of components: central narrow emission of nebular type, broad emission component indicating optically thin stellar wind with a high velocity of $\sim \!\! 500$~km\,s$^{-1}$, very broad multicomponent P~Cyg absorption, and satellite emission components situated on both sides of the central peak of the line with a velocity of more than $1000$~km\,s$^{-1}$, indicating bipolar collimated outflow. Both satellite components were observed only during the time when the intensity of the H$\alpha$ satellite components was maximal.

The profile of the H$\alpha$ and H$\beta$ lines of the Hen~3-1341 system during its 1998--2004 outburst had high velocity satellite emission components on both sides of the central peak of the line with a velocity of $\sim\!\!800$~km\,s$^{-1}$, indicating bipolar collimated outflow from the outbursting compact object. The same components presented in the profile of the helium triplet line with wavelength $\lambda$\,5876, but this line contained broad two-component P~Cyg absorption in addition \citep{tomata00,Munari+05}.

The profile of the H$\beta$ line of the BF~Cyg system during its 2006--2012 outburst contained satellite components with a velocity of $\sim\!\!400$~km\,s$^{-1}$ situated on both sides of the line center indicating bipolar collimated outflow from the outbursting compact object. The blueshifted satellite component was an absorption one at the more early stage of the outburst which after that went into emission. At the time when the blueshifted component appeared as absorption, an other absorption of P~Cyg type at a velocity position of $250$~km\,s$^{-1}$ presented in the spectrum. A second P~Cyg absorption with lower velocity of $50$--$100$~km\,s$^{-1}$ was seen at same time. The H$\alpha$ line had weak emission peaks/shoulders with the same velocity of $\sim\!\!400$~km\,s$^{-1}$ situated on both sides of the central emission like the H$\beta$ satellite components.

The line profiles of the Z~And, Hen~3-1341, and BF~Cyg systems are considered in the light of the model about the second stage of the evolution of the accreting compact object which stage is related to the strong requrrent outbursts of Z~And. It is supposed that a geometrically thick disk-like formation surrounding the compact object exists in the system which plays a role of mechanism of collimation. The high-velocity stellar wind collides with this formation and collimated outflow forms after the collision. The region of the wind which is projected onto the effective photosphere (pseudophotosphere) of the outbursting compact object gives rise to the P~Cyg absorption.

The spectral lines of the symbiotic binary AG~Dra did not contain indication of collimated outflow. The \mbox{He\,{\sc ii}} $\lambda$\,4686 line had a broad emission component in whose appearance three processes can take place: stellar wind, rotation of an accretion disc, and electron scattering. We supposed that the stellar wind is possible to collide with the disc but this collision does not lead to the appearance of optically thick shell occulting the outbursting compact object as in the Z~And system.

\begin{theacknowledgments}

The authors are grateful to Dr.~T.~Tomov for the kindly given opportunity to use the spectrum of Hen~3-1341 in the region of the \mbox{He\,{\sc i}} line. This work has been supported by the Bulgarian Scientific Research Fund (Grants 01/14 from BSTC  Bulgaria--Slovakia and DO~02-85), the Basic Research Program of the Presidium of the Russian Academy of Sciences, Russian Foundation for Basic Research (projects 11-02-00076 and 11-02-01248), Federal Targeted Program ``Science and Science Education for Innovation in Russia 2009--2013,'' and the Russian and Bulgarian Academies of Sciences through a collaborative program in basic space research.

\end{theacknowledgments}

\end{document}